\documentclass[a4paper]{article}

\usepackage{amsmath}
\usepackage{amsfonts}
\usepackage{amssymb}         
\usepackage{amsopn}
\usepackage{theorem}          
\usepackage{latexsym}
\usepackage{graphicx}        
\usepackage{mathrsfs}		
\usepackage{psfrag}
\usepackage{algorithmic}
\usepackage{algorithm}
\usepackage{url}

\newtheorem{result}{\ }[section]
\theoremstyle{changebreak}                

\newtheorem{lem}[result]{Lemma}

\newtheorem{eg}[result]{Example}
\newenvironment{proof}
 {{\sl Proof.}\hspace*{1 ex}}%
 {{\nopagebreak\hspace*{\fill}$\Box$\par\vspace{12pt}}}
\begin{document}

\thispagestyle{empty}
\begin{center} 

{\LARGE Fast paths in large-scale dynamic road networks} 
\par \bigskip
{\sc Giacomo Nannicini${}^{1,2}$, Philippe Baptiste${}^1$, Gilles
  Barbier${}^{2}$, Daniel Krob${}^1$, Leo Liberti${}^1$}   
\par \bigskip
\begin{minipage}{15cm}
\begin{flushleft}
{\small
\begin{itemize}
\item[${}^1$] {\it LIX, \'Ecole Polytechnique, F-91128 Palaiseau,
  France} \\ Email:\url{{giacomon,baptiste,dk,liberti}@lix.polytechnique.fr}
\item[${}^2$] {\it Mediamobile, 10 rue d'Oradour sur Glane, Paris,
  France} \\ Email:\url{{giacomo.nannicini,gilles.barbier,contact}@v-trafic.com}
\end{itemize}
}
\end{flushleft}
\end{minipage}
\par \medskip \today
\end{center}
\par \bigskip

\begin{abstract} 
Efficiently computing fast paths in large-scale dynamic road networks
(where dynamic traffic information is known over a part of the
network) is a practical problem faced by several traffic information
service providers who wish to offer a realistic fast path computation
to GPS terminal enabled vehicles. The heuristic solution method we
propose is based on a highway hierarchy-based shortest path algorithm
for static large-scale networks; we maintain a static highway
hierarchy and perform each query on the dynamically evaluated network.
\end{abstract}

\section{Introduction}
\label{introduction}
Several cars are now fitted with a Global Positioning System (GPS)
terminal which gives the exact geographic location of the vehicle on
the surface of the earth. All of these GPS terminals are now endowed
with detailed road network databases which allow them to compute the
shortest path (in terms of distance) between the current vehicle
location (source) and another location given by the driver
(destination). Naturally, drivers are more interested in the
source-destination {\it fastest} path (i.e. shortest in terms of
travelling time). The greatest difficulty to overcome is that the
travelling time depends heavily on the amount of traffic on the chosen
road. Currently, some state agencies as well as commercial enterprises
are charged with monitoring the traffic situation in certain
pre-determined strategic places. Furthermore, traffic reports are
collected from police cars as well as some taxi services. The dynamic
traffic information, however, is as yet limited to a small proportion
of the whole road network.

The problem faced by traffic information providers is currently that
of offering GPS terminal enabled drivers a source-destination path
subject to the following constraints: (a) the path should be fast in
terms of travelling time subject to dynamic traffic information being
available on part of the road network; (b) traffic information data
are updated approximately each minute; (c) answers to path queries
should be computed in real time. Given the data communication time and
other overheads, constraint (c) practically asks for a shortest path
computation time of no more than 1 second. Constraint (b) poses a serious
problem, because it implies that the fastest source-destination path
may change each minute, giving an on-line dimension to the problem. A
source-destination query spanning several hundred kilometers, which
would take several hours to travel, would need a system recomputing
the fastest path each minute; this in turn would mean keeping track of
each query for potentially several hours. As the estimated
computational cost of this requirement is superior to the resources
usually devoted to the task, a system based on dynamic traffic
information will not, in practice, ever compute the on-line fastest
path. As a typical national road network for a large European country
usually counts several million junctions and road segments, constraint
(c) implies that a straight Dijkstra's algorithm is not a viable
option. In view of constraint (a), in our solution method fast paths
can be efficiently computed by means of a point-to-point
hierarchy-based shortest path algorithm for static large-scale
networks, where the hierarchy is built using static information and
each query is answered on the dynamically evaluated network.

This paper makes two original scientific contributions (i) We extend a known hierarchy-based shortest
path algorithm for static large-scale undirected graphs (the Highway Hierarchies
algorithm \cite{schultes}) to the directed case. The method has been
developed and tested on real road network data taken from the
TeleAtlas France database \cite{teleatlas}. We note that the
original authors of \cite{schultes} have extended the algorithm to
work on directed graphs in a slightly different way than ours (see \cite{schultesdir}).
(ii) We propose a method for efficiently finding fast paths on a
large-scale dynamic road network where arc travelling times are
updated in quasi real-time (meaning very often but not continuously).

In the rest of this section, we discuss the state of the art as
regards shortest path algorithms in dynamic and large-scale networks,
and we describe the proposed solution. The rest of the paper is
organized as follows. In Section \ref{schultes} we briefly review the highway
hierarchy-based shortest path algorithm for static large-scale
networks, which is one of the important building blocks of our method,
and discuss the extension of the existing shortest-path algorithm to
the directed case. Section \ref{compres} discusses the computational
results, and Section \ref{conclusion} concludes the paper.

\subsection{Shortest path algorithms in road networks}
\label{litrev}
The problem of computing fastest paths in graphs whose arc weights
change over time is termed the {\sc Dynamic Shortest Path Problem}
(DSPP) \cite{brt}. The work that laid the foundations for solving the
DSPP is \cite{cookehalsey} (a good review of this paper can be found
in \cite{dreyfus}, p.~407): Dijkstra's algorithm is extended to the
dynamic case through a recursion formula based on the assumption that
the network $G=(V,A)$ has the FIFO property: for each pair of time
instants $t,t'$ with $t<t'$:
\begin{equation*}
  \forall\; (u,v)\in A \ \tau_{uv}(t) + t \le \tau_{uv}(t') + t',
\end{equation*}
where $\tau_{uv}(t)$ is the travelling time on the arc $(u,v)$
starting from $u$ at time $t$. The FIFO property is also called the
{\it non-overtaking property}, because it basically says that if $A$
leaves $u$ at time $t$ and $B$ at time $t'>t$, $B$ cannot arrive at
$v$ before $A$ using the arc $(u,v)$. The shortest path problem in
dynamic FIFO networks is therefore polynomially solvable
\cite{chabini1}, even in the presence of traffic lights \cite{aops}.
Dijkstra's algorithm applied to dynamic FIFO networks has been
optimized in various ways \cite{brt,chabini1}; the $A^\ast$ one-to-one
shortest path algorithm has also been extended to dynamic networks
\cite{chabini2}. The DSPP is NP-hard in non-FIFO networks
\cite{dean1}.

Although in this paper we do not assume any knowledge about the
statistical distribution of the arc weights in time, it is worth
mentioning that a considerable amount of work has been carried out for
computing shortest paths in stochastic networks. A good review is
\cite{sintef}.

The computation of exact shortest paths in large-scale static networks
has received a good deal of attention \cite{chanzhang}. The
established practice is to delegate a considerable amount of
computation to a preprocessing phase (which may be very slow) and then
perform fast source-destination shortest path queries on the
pre-processed data. Recently, the concept of {\it highway hierarchy}
was proposed in \cite{schultes,schultesmaster,schultesdir}. A highway hierarchy of
$L$ levels of a graph $G=(V,A)$ is a sequence of graphs
$G=G^0,\ldots,G^{L}$ with vertex sets $V^0 = V, V^1 \supseteq \ldots \supseteq V^L$ and arc sets
$A^0 = A, A^1 \supseteq \ldots \supseteq A^L$; each arc has maximum hierarchy level
(the maximum $i$ such that it belongs to $A^i$) such that for all pairs of
vertices there exists between them a shortest path $(a_1,\ldots,a_k)$, where $a_i$
are the consecutive path arcs, whose search level
first increases and then decreases, and each arc's search level is not greater
than its maximum hierarchy level. A more precise description is given in Section
\ref{schultes}. The $A^\ast$ algorithm has also been extended to use
a concept, {\it reach}, which has turned out to be closely related to highway hierarchies
(see \cite{goldberg}).

\subsection{Description of the solution method}
\label{secmainalgs}
The solution method we propose in this paper efficiently finds fast
paths by deploying Dijkstra-like queries on a highway hierarchy built
using the static arc weights found in the road network database, but
used with the dynamic arc weights reflecting quasi real-time traffic
observations. This
implies using two main building blocks: highway
hierarchy construction (the Highway Hierarchies\footnote{From now on,
simply HH} algorithm extended to directed graphs),
and the query algorithm. Consequently, the implementation is
a complex piece of software whose architecture has been detailed in
the appendix.
\begin{itemize}
\item {\it Highway hierarchy}. Apply the directed graph extension of
  the HH algorithm (see Section \ref{schultes}) to construct a
  highway hierarchy using the static road network information. In
  particular, arc travelling times are average estimations found in
  the database. This is a preprocessing step that has to be performed
  only when the topology of the road network changes. The CPU time
  taken for this step is not an issue.
\item {\it Efficient path queries}. Efficiently address
  source-destination fast path requests by employing a multi-level
  bidirectional Dijkstra's algorithm on the dynamically evaluated
  graph using the highway hierarchy structure constructed during
  preprocessing. This algorithm is carried out each time a path
  request is issued; its running time must be as fast as possible, in
  any case not over 1 second.
\end{itemize}

\section{Highway Hierarchies algorithm on dynamic directed graphs} 
\label{schultes}
The Highway Hierarchies algorithm \cite{schultes,schultesmaster} is a fast,
hierarchy-based shortest paths algorithm which works on static undirected
graphs. HH algorithm is specially suited to efficiently
finding shortest paths in large-scale networks. Since the HH
algorithm is one of our main building blocks, we briefly review the
necessary concepts.

The Highway Hierarchies algorithm is heavily based on Dijkstra's algorithm
\cite{dijkstra}, which finds the tree of all shortest paths from a
root vertex $r$ to all other vertices $v\in V$ of a weighted digraph
$G=(V,A)$ by maintaining a heap $H$ of {\it reached} vertices $u$ with
their associated (current) shortest path length $c(u)$ (elements of
the heap are denoted by $[u,c(u)]$. Vertices which have not yet
entered the heap (i.e.~which are still unvisited) are {\it unreached},
and vertices which have already exited the heap (i.e.~for which a
shortest path has already been found) are {\it settled}. Dijkstra's
algorithm is as follows.
\begin{enumerate}
\item Let $H=\{[r,0]\}$.
\item If $H=\emptyset$, terminate.
\item Let $u$ be the vertex in $H$ with minimum associated path length
  $c(u)$.
\item Let $H=H\smallsetminus \{u\}$.
\item For all $v\in\delta^+(u)$, if $c(u)+\tau_{uv}<c(v)$ then let
  $H= (H\smallsetminus \{[v,c(v)]\}) \cup \{[v,c(u)+\tau_{uv}]\}$.
\item Go to 2.
\end{enumerate}
A {\it bidirectional} Dijkstra algorithm works by keeping track of two
Dijkstra search scopes: one from the source, and one from the
destination working on the reverse graph. When the two search scopes
meet it can be shown that the shortest path passes through a vertex
that has been reached from both nodes (\cite{schultesmaster},
p.~30). A set of shortest paths is {\it canonical}\footnote{Dijkstra's
algorithm can easily be modified to output a canonical shortest paths
tree (see \cite{schultesmaster}, Appendix A.1 --- can be downloaded
from \url{http://algo2.iti.uka.de/schultes/hwy/}).} if, for
any shortest path $p=\langle
u_1,\ldots,u_i,\ldots,u_j\ldots,u_k\rangle$ in the set, the canonical
shortest path between $u_i$ and $u_j$ is a subpath of $p$.

The HH algorithm works in two stages: a time-consuming
pre-processing stage to be carried out only once, and a fast query
stage to be executed at each shortest path request. Let
$G^0=G$. During the first stage, a highway hierarchy is constructed,
where each hierarchy level $G^l$, for $l\le L$, is a modified subgraph
of the previous level graph $G^{l-1}$ such that no canonical shortest
path in $G^{l-1}$ lies entirely outside the current level for all
sufficiently distant path endpoints: this ensures that all queries
between far endpoints on level $l-1$ are mostly carried out on level
$l$, which is smaller, thus speeding up the search. Each shortest path
query is executed by a multi-level bidirectional Dijkstra algorithm:
two searches are started from the source and
from the destination, and the query is completed shortly after the
search scopes have met; at no time do the search scopes decrease
hierarchical level. Intuitively, path optimality is due to the fact
that by hierarchy construction there exist no canonical shortest path
of the form $\langle a_1,\ldots,a_i,\ldots,a_j\ldots,a_k\ldots\rangle$, where
$a_i,a_j,a_k\in A$ and the search level of $a_j$ is lower than the level of
both $a_i,a_k$; besides, each arc's search level is always lower or equal to
that arc's maximum level, which is computed during the hierarchy construction
phase and is equal to the maximum level $l$ such that the arc belongs to $G^l$.
The speed of the query is due to the fact that the
search scopes occur mostly on a high hierarchy level, with fewer arcs and
nodes than in the original graph.

\subsection{Highway hierarchy}
\label{hierarchy}
As the construction of the highway hierarchy is the most complicated
part of HH algorithm, we endeavour to explain its main traits
in more detail. Given a local extensionality parameter $H$ (which
measures the degree at which shortest path queries are satisfied
without stepping up hierarchical levels) and the maximum number of
hierarchy levels $L$, the iterative method to build the next highway
level $l+1$ starting from a given level graph $G^l$ is as follows:
\begin{enumerate}
\item For each $v\in V$, build the neighbourhood $N_H^l(v)$ of all
  vertices reached from $v$ with a simple Dijkstra search in the
  $l$-th level graph up to and including the $H$-st settled
  vertex. This defines the local extensionality of each vertex,
  i.e. the extent to which the query ``stays on level $l$''.
\item For each $v\in V$:
\begin{enumerate}
\item Build a partial shortest path tree $B(v)$ from $v$, assigning a
  status to each vertex. The initial status for $v$ is ``active''. The
  vertex status is inherited from the parent vertex whenever a vertex
  is reached or settled. A vertex $w$ which is settled on the shortest
  path $\langle v,u,\ldots,w\rangle$ (where $v\not= u\not= w$)
  becomes ``passive'' if  \label{psptree}
\begin{equation}
  |N_H^l(u)\cap N_H^l(w)|\le 1. 
  \label{passivity}
\end{equation}
  The partial shortest path tree is complete when there are no more
  active reached but unsettled vertices left.
\item From each leaf $t$ of $B(v)$, iterate backwards along the branch
  from $t$ to $v$: all arcs $(u,w)$ such that $u\not\in N_H^l(t)$ and
  $w\not\in N_H^l(v)$, as well as their adjacent vertices $u,w$, are
  raised to the next hierarchy level $l+1$. \label{lifthigher}
\end{enumerate}
\item Select a set of {\it bypassable} nodes on level $l+1$; intuitively, these
  nodes have low degree, so that the benefit of skipping them during a search
  outweights the drawbacks (i.e., the fact that we have to add shortcuts to
  preserve the algorithm's correctness). Specifically, for a given set
  $B_{l+1} \subset V_{l+1}$ of bypassable nodes, we define
  the set $S_{l+1}$ of shortcut edges that bypass the nodes in $B_{l+1}$: for each path $p =
  (s, b_1, b_2,\ldots, b_k, t)$ with $s, t \in V_{l+1} \smallsetminus Bl$ and
  $b_i \in B_{l+1}, 1\le i \le k$, the set $S_{l+1}$ contains
  an edge $(s, t)$ with $c(s, t) = c(p)$. The core $G'_{l+1} = (V'_{l+1}, E'_{l+1})$
  of level $l+1$ is defined as:$V'_{l+1} = V_{l+1} \smallsetminus B_{l+1}$,
  $E'_{l+1} = (E_{l+1} \cap (V'_{l+1} \times V'_{l+1})) \cup S_{l+1}$.
\end{enumerate}
The result of the contraction is the contracted highway network $G'_{l+1}$,
which can be used as input for the following iteration of the construction procedure.
It is worth noting that higher level graphs may be disconnected even
though the original graph is connected.

\begin{eg}
\label{ex1}
Take the directed graph $G=(V,A)$ given in Fig.~\ref{ex1fig}
(above). We are going to construct a road hierarchy with $H=3$ and
$L=1$ on $G$. First we compute $N_3^0(v)$ for all $v\in
V=\{v_0,\ldots,v_6\}$.
\begin{center}
\begin{tabular}{l|l}
$v$ & $N_3^0(v)$ \\ \hline
$v_0,v_1,v_2$ & $\{0,1,2\}$  \\
$v_3,v_4,v_5$ & $\{3,4,5\}$  \\
$v_6$ & $\{3,5,6\}$ 
\end{tabular}
\end{center}
Next, we compute $B(v)$ for all $v\in V$ and raise the hierarchy level
of the relevant arcs from the leaves to $B(v)$ to $v$. We only discuss
the computation of $B(v_0)$ in detail as the others are
similar. 
\begin{enumerate}
\item Vertex $v_0$ is initialized as an active vertex. 
\item Dijkstra's algorithm is started. 
\begin{enumerate}
\item $v_0$ is settled (cost $0$) on the empty path, so the passivity
  condition (\ref{passivity}) does not apply; 
\item $v_1$ and $v_2$ are reached from $v_0$ with costs resp.~$1$ and
  $2$, and inherit its active status;
\item $v_1$ is settled (cost $1$) on the path $\langle v_0,v_1\rangle$
  and condition (\ref{passivity}) does not apply; 
\item $v_6$ is reached from $v_1$ with cost $1+4=5$ and set to
  active;
\item $v_2$ (cost $2$) is settled on $\langle v_0,v_2\rangle$; 
\item $v_4$ is reached from $2$ with cost $2+6=8$ and set to active;
\item $v_6$ (cost $5$) is settled on the path $\langle
  v_0,v_1,v_6\rangle$: since $N_3^0(v_1)\cap N_3^0(v_6) = \emptyset$,
  condition (\ref{passivity}) is verified, and $v_6$ is labeled
  passive; 
\item $v_3$ is reached from $v_6$ with cost $1+4+4=9$ and set to
  passive. 
\item $v_4$ (cost $8$) is settled on the path $\langle
  v_0,v_2,v_4\rangle$: since $N_3^0(v_2)\cap N_3^0(v_4) = \emptyset$,
  condition (\ref{passivity}) is verified, and $v_4$ is labeled
  passive; 
\item $v_5$ is reached from $v_4$ with cost $2+6+2=10$ and set to
  passive;
\item the only unsettled vertices are $v_3$ and $v_5$. Since 
both are reached and passive, the search terminates. 
\end{enumerate}
\item The leaf vertices of $B(v_0)$ are $v_4$ and $v_6$. 
\begin{enumerate}
\item From $t=v_4$, we iterate backwards along the arcs on the path
  $\langle v_0,v_2,v_4\rangle$: the arc $(v_2,v_4)$ has the property
  that $v_2\not\in N_3^0(v_4)$ and $v_4\not\in N_3^0(v_2)$, so its
  hierarchy level is raised to $l+1=1$ (the other arc on the path,
  $(v_0,v_2)$, stays at level $l=0$);
\item from $t=v_6$, we iterate backwards along the arcs on the path
$\langle v_0,v_1,v_6\rangle$: the arc $(v_1,v_6)$ has the property
that $v_1\not\in N_3^0(v_6)$ and $v_6\not\in N_3^0(v_1)$, so its
hierarchy level is raised to $1$ (the other arc on the path stays at
level $0$).
\end{enumerate}
\end{enumerate}
Fig.~\ref{ex1fig} shows the hierarchy at level 1.
\begin{figure}[!ht]
\begin{center}
\fbox{
\includegraphics[width=7cm, height=5cm]{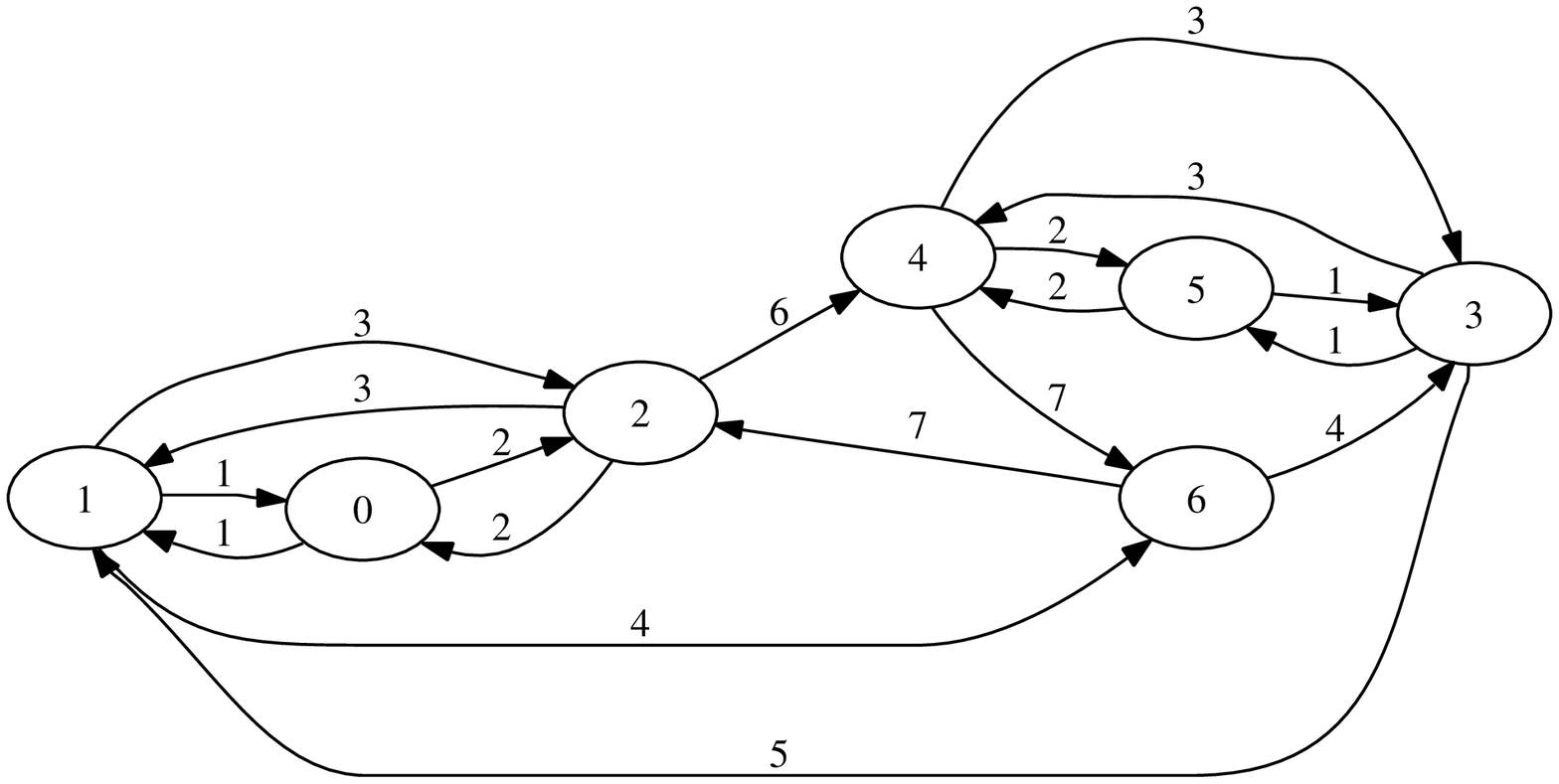}
\includegraphics[width=7cm, height=5cm]{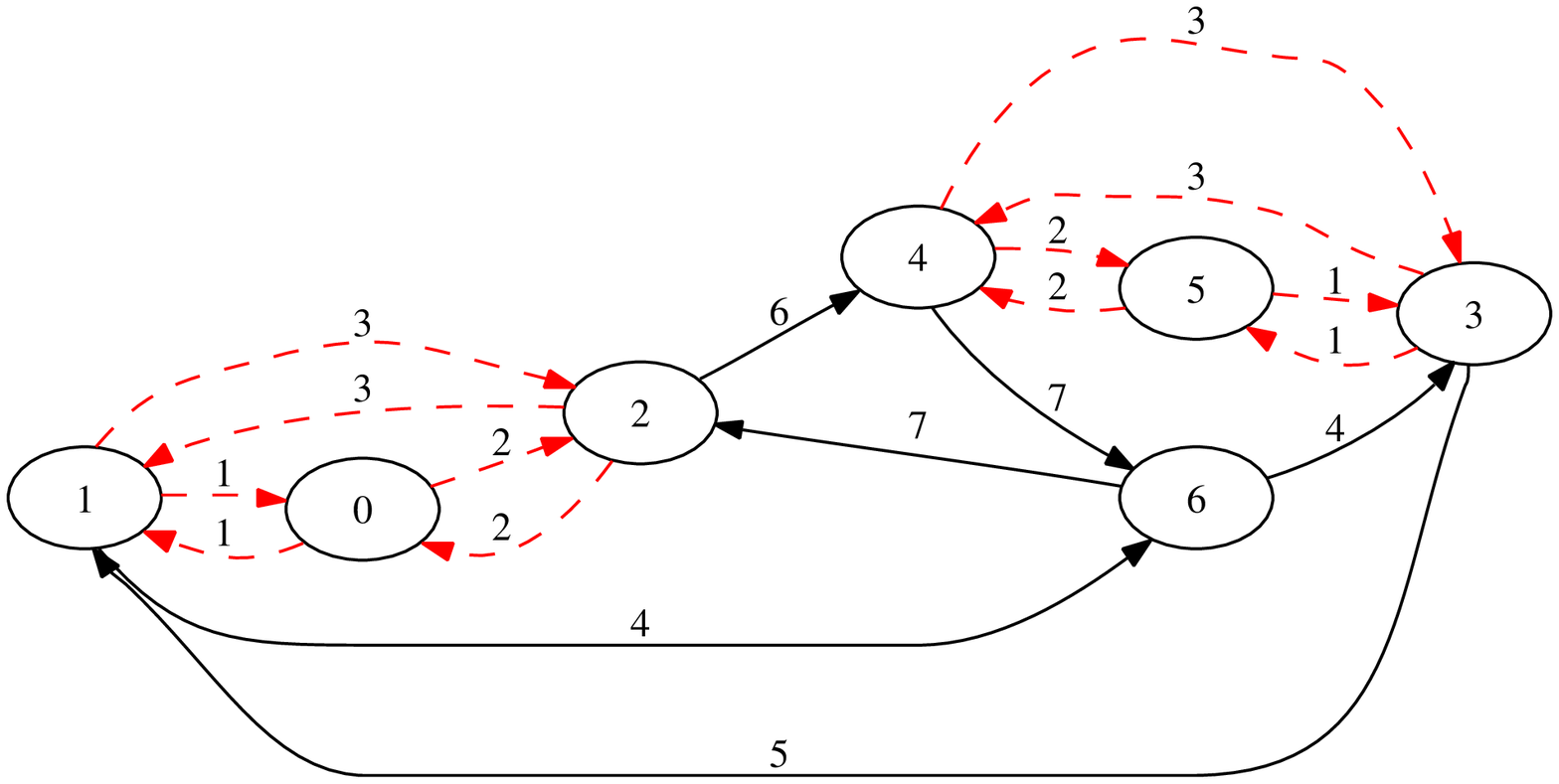}
}
\end{center}
\caption{The graph of Example \ref{ex1} (left) and its highway
hierarchy for $H=3,L=1$ (right): the dashed lines indicate arcs at
level 0, the solid lines indicate arcs at level 1.}
\label{ex1fig}
\end{figure}
\end{eg}

\subsection{Extension to directed graphs}
The original description of the HH algorithm \cite{schultes}
applies to undirected graphs only; in this section we provide an extension to the directed case.
It should be noted that the HH algorithm was extended to the directed case by the authors (see \cite{schultesdir}) in a way which is very similar to that described here. 
However, we believe our slightly different exposition helps to clarify these ideas considerably.

The algorithm for hierarchy construction, as explained in
Section~\ref{hierarchy}, works with both undirected and directed
graphs. However, storing all neighbourhoods $N_H^l(v)$ for each $v$
and $l$ has prohibitive memory requirements. Thus, the original HH
implementation for checking whether a vertex $v$ is in $N_H^l(u)$ is
based on comparing the distance $d(u,v)$ with the
``distance-to-border'' (also called {\it slack}) from $u$ to the
border of its neighbourhood $N_H^l(u)$. The ``distance-to-border'' $d_H^l(u)$
is a measure of a neighbourhood's radius, and is defined as the distance
$d(u,v)$ where $v$ is the farthest node in $N_H^l(u)$, i.e. the cost of the
shortest path from $u$ to the $H$-th settled node when applying Dijkstra's algorithm
on node $u$ at level $l$. This is the basis of the
slack-based method in \cite{schultesmaster}, p.~19 (from which we draw
our notation). In the partial shortest paths tree $B(s_0)$ computed in
Step \ref{psptree} of the algorithm in Section \ref{hierarchy}, the
slack $\Delta(u)$ is recursively computed for all $u\in B(s_0)$
starting from the leaves $t_0$ of $B(s_0)$, as follows.
\begin{enumerate}
\item Initialise a FIFO queue $Q$ to contain all nodes $u$ of
$B(s_0)$, ordering them from the farthest one to the nearest one with respect to $s_0$.
\item Set $\Delta(u)=d_H^l(u)$ for $u$ a leaf of $B(s_0)$ and
$+\infty$ otherwise. \label{slackinit}
\item If $Q$ is empty, terminate. \label{goto1}
\item Remove $u$ from $Q$, and let $p$ be its predecessor in
  $B(s_0)$. 
\item If $\Delta(p)=+\infty$ and $p\not\in N_H^l(s_0)$, $p$ is added
  to $Q$.
\item Let $\Delta(p)=\min(\Delta(p),\Delta(u)-d(p,u))$.
\item If $\Delta(p)<0$, the edge $(p,u)$ is lifted to the higher
  hierarchical level.
\item Return to Step \ref{goto1}.
\end{enumerate}
The algorithm works because Thm.~2 in \cite{schultesmaster} proves
that condition $\Delta(p)<0$ is equivalent to the condition of Step
\ref{lifthigher} of the algorithm in Section \ref{hierarchy}. The
cited theorem is based on the following assumption:
\begin{equation}
  \forall u\in V\; (u\not\in N_H^l(t_0) \rightarrow
    d_H^l(t_0)-d(u,t_0)<0).
  \label{wrongass}
\end{equation}
This condition may fail to hold for directed graphs, since $d(u,t_0)
\neq d(t_0,u)$.

\begin{figure}[!htb]
\begin{center}
\fbox{\includegraphics[width=11cm]{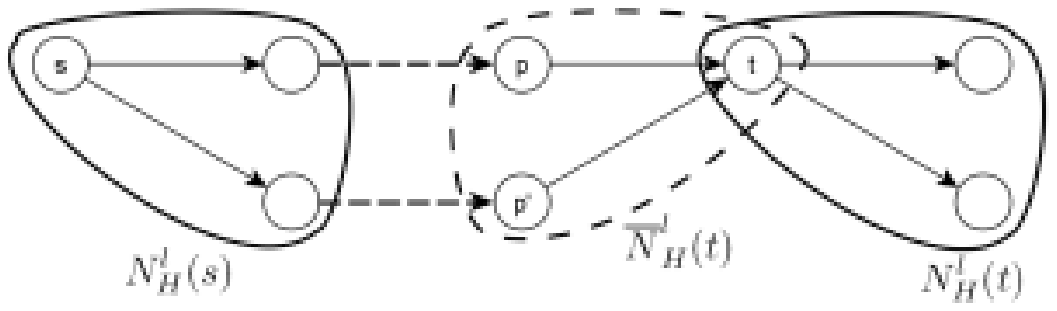}}
\end{center}
\caption{An example which shows neighbourhoods and reverse
  neighbourhoods with $H=3$; only solid arcs are lifted to a higher
  level in the hierarchy. Note that arcs $(p,t)$ and $(p',t)$ are not
  lifted even if $p,p' \notin N_H^l(t)$; this is because $p,p' \in
  \overline{N}_H^l(t)$, and for target node we consider the reverse
  neighbourhood. }
\label{inv_neigh}
\end{figure}

To make Assumption \ref{wrongass} hold, we have to consider a
neighbourhood radius computed on the reverse graph, that is the graph
$\overline{G} = (V,\overline{A})$ such that $(u,v) \in \overline{A}
\Leftrightarrow (v,u) \in A$. Thus, we modified the original
implementation to compute, for each node, a reverse neighbourhood
$\overline{N}_H^l(v)$ (see Figure \ref{inv_neigh}), so that we are
able to store the corresponding reverse neighbourhood radius
$\overline{d}_H^l(u) \, \forall u \in V$.
We replace Step \ref{slackinit} in the algorithm above by:
\begin{quote}
  \ref{slackinit}a. Set $\Delta(u)=\overline{d}_H^l(u)$ for $u$ a leaf
  of $B(s_0)$ and $+\infty$ otherwise. \label{slackinita}
\end{quote}
We are now going to prove our key lemma.
\begin{lem}
\label{keylemma}
Let $u,s\in V$ and $t$ a
leaf in $B(s)$. If $u\not\in \overline{N}_H^l(t)$ then
$\overline{d}_H^l(t)-d(u,t)<0$.
\end{lem}
\begin{proof}
Suppose $d(u,t)\le \overline{d}_H^l(t)$. By definition, this means that there
is a shortest path in $\overline{N}_H^l(t)$ which connects $u$ to
$t$. Therefore, $u\in \overline{N}_H^l(t)$ against the hypothesis. 
\end{proof}
It is now straightforward to amend Thm.~2 in \cite{schultesmaster} to
hold in the directed case; all other theorems in \cite{schultesmaster}
need similar modifications, replacing $N_H^l(t)$ with
$\overline{N}_H^l(t)$ and $d_H^l(t)$ with $\overline{d}_H^l(t)$
whenever $t$ is target node or is ``on the right side'' of a path - it
will always be clear from the context. The query algorithm must me
modified to cope with these differences, using $\overline{d}_H^l(t)$
instead of $d_H^l(t)$ whenever we are searching in the backwards
direction.

Interestingly, the problem with the slack-based method was first detected when our original
implementation of the HH algorithm failed to construct a correct
hierarchy for the Paris urban area. This shows that the
extension of the algorithm to the directed case actually arises from
real needs.

\subsection{Heuristic application to dynamic networks}
The original Highway Hierarchies algorithm, as described above, finds shortest paths in
networks whose arc weights do not change in time. By forsaking the
optimality guarantee, we adapt the algorithm to the case of networks
whose arc weights are updated in quasi real-time. Whereas the highway
hierarchy is constructed using the static arc travelling times from
the road network database, each point-to-point path query is deployed
on a dynamically evaluated version of the highway hierarchy where the
arcs are weighted using the quasi real-time traffic information. In
particular, in all tests that involved a comparison with neighbourhood
radius we use the static arc travelling times, while for all
evaluations of path lengths or of node distances we use the real-time
(dynamic) travelling times. This means that the static travelling
times are used to determine neighbourhood's crossings, and thus to
determine when to switch to a higher level in the hierarchy, while the
key for the priority queue for HH algorithm is computed using only
dynamic travelling times.

\section{Computational results}
\label{compres}
In this section we discuss the computational results obtained by our
implementation. As there seems to be no other readily available software
with equivalent functionality, the computational results are not comparative. However, we
establish the quality of the heuristic solutions by comparing them
against the fastest paths found by a plain Dijkstra's algorithm.
We mention here two different approaches: dynamic highway-node routing 
(\cite{schultesdyn}), which uses a selection of nodes operated by the HH algorithm
to build an overlay graph (see \cite{wagnerps}), and dynamic ALT (\cite{dynALT}), which is a dynamic
landmark-based implementation of $A^\ast$. Both approaches, however, although
very performing with respect to query times, require a computationally heavy update phase
(which takes time in the order of minutes), and thus are not suitable for our scenario,
where, supposedly, each arc can have its cost changed every 2 minutes (roughly).

We performed the tests on the entire road network of France, using a
highway hierarchy with $H=65$ and $L=9$. The original network has $7778913$
junctions and $17154042$ road segments; the number of nodes and arcs in each level is as
follows.
\begin{center}
\begin{tabular}{l||l|l|l|l|l|l|l|l|l|l}
{\bf level} & 0 & 1 & 2 & 3 & 4 & 5 & 6 & 7 & 8 & 9 \\ \hline
{\bf nodes}  & 7778913 & 1517291 & 433286 & 182474 & 91888 & 53376 & 34116 & 23338 & 16445 & 11790 \\
{\bf arcs}  & 17154042 & 3461385 & 1283000 & 583380 & 308249 & 183659 & 119524 & 81170 & 57235 & 41092  \\
\end{tabular}
\end{center}

We show the results for queries on the full graph without dynamic
 travelling times in Table \ref{tabcomp}; in this case, all paths
 computed with the HH algorithm are fastest paths. In Table
 \ref{tabcomp2}, instead, we record our results on a graph with
 dynamic travelling times; we also report the relative distance of the
 solution found with our heuristic version of the HH algorithm and the
 fastest path computed with Dijkstra, and, for comparative reasons,
 the results of the naive approach which consists in computing the
 traffic-free optimal solution with the HH algorithm (i.e., on the
 static graph) and then applying dynamic times on the so-found
 solution. Dynamic travelling times were taken choosing, for each
 query, one out of five sets of values recorded in different times of
 the day for each of the $29384$ arcs with dynamic information.

Although this number is
 small with respect to the total number of arcs in the graph, it
 should be noted that most of these arcs correspond to very important
 road segments (highways and national roads). All arcs $(i,j)$ that
 did not have a dynamic travelling time were assigned a different
 weight at each query, chosen at random with a uniform distribution
 over $[\tau_{ij},15\tau_{ij}]$, where $\tau_{ij}$ is the reference
 time for arc $(i,j)$. This choice has been made in order to recreate
 a difficult scenario for the query algorithm: even if the number of
 arcs with real traffic information is still small, it is going to
 increase rapidly as the means for obtaining dynamic information
 increase (e.g.~number of road cameras, etc.), and thus, to simulate
 an instance where most arcs have their travelling time changed
 several times per hour, we generated each arc's cost at random. The
 interval $[\tau_{ij},15\tau_{ij}]$ is simply a rough estimation of a
 likely cost interval, based on the analysis of historical data.
All tables report average values over 5000 queries. All computational
 results in Table \ref{tabcomp} and \ref{tabcomp2} have been obtained
 on a multiprocessor Intel Xeon 2.6 GHz with 8GB RAM running Microsoft
 Windows Server 2003, compiling with Miscrosoft Visual Studio 2005 and
 optimization level 2.

Computational results show that, although with no guarantee of
optimality, our heuristic version of the algorithm works well in
practice, with $0.55\%$ average deviation from the optimal solution
and a recorded maximum deviation of $17.59\%$; query times do not seem
to be influenced by our changes with respect to the original version
of the algorithm. The naive approach of computing the shortest path on
the static graph, and then applying dynamic times, records an average
error of $2\%$, but it has a much higher variance, and a maximum error
of $27.95\%$; although the average error is not high, it's still
almost $4$ times the average error of the more sofisticated approach,
and the high variance suggests lack of stability in the solution's
quality. The low value recorded for the average error with the naive
approach (in absolute terms) can be explained as a consequence of the
following two facts: travelling times generated at random on arcs
without real-time traffic information cannot simulate real traffic
situation, because they lack spatial coherence (i.e. they do not
simulate congested nearby zones) and traffic behaviour information
(i.e. the fact that during peak hours important road segments are
likely to be congested, while less important roads are not), thus
making the task of finding a fast path easier; besides, the average
query on such a large graph corresponds to a very long path (296
minutes on the traffic-free graph, 2356 minutes on the dynamic graph),
and on long paths it is usually necessary to use highways or national
roads regardless of their congestion status - which is exactly what
the HH algorithm does. This last sentence is supported by the fact
that, if we consider only the $500$ shortest queries in terms of path
length, the average error of the naive approach increases to $3.60\%$,
while the average error of the heuristic version increases to
$0.97\%$; this is in accord with the fact that on short paths the
influence of traffic is greater, because alternative routes that do
not use highways are more appealing, while on long paths using
highways is often a necessary step.
However, in relative terms, the heuristic version of the HH algorithm
performs significantly better than the naive approach proposed for
comparison, and we expect the difference to increase (in favour of the
heuristic algorithm) if applied to a graph fully covered with real
traffic information.

\begin{figure}[!ht]
\begin{center}
\fbox{\includegraphics[width=15cm]{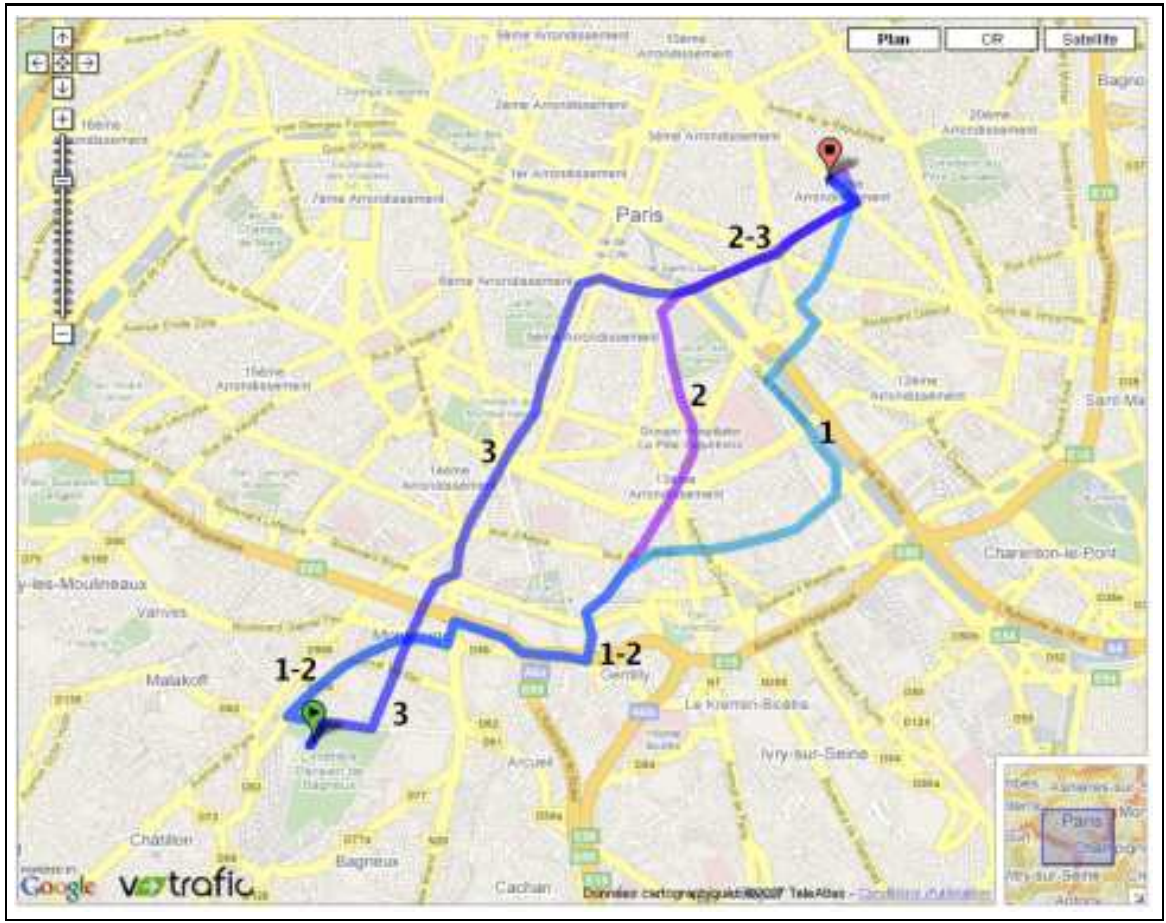}}
\end{center}
\caption{Fast paths calculated with different algorithms; each number
  identifies a path, paths are partially overlapping. \textbf{1}:
  Dijkstra's algorithm (optimal solution) with real-time arc costs;
  dynamic travelling time: 24 minutes and 6 seconds. \textbf{2}: HH
  algorithm (heuristic solution) with real-time arc costs; dynamic
  travelling time: 25 minutes and 5 seconds. \textbf{3}: HH algorithm
  without real-time arc costs (traffic-free optimal solution); dynamic
  travelling time: 37 minutes and 5 seconds.}
\label{paths}
\end{figure}

Figure \ref{paths} shows how the optimal and the heuristic path may
differ; since the hierarchy built on the static graph emphasizes
important roads, the heuristic algorithm applied on the dynamically
weighted graph still tends to use highways and national roads even
when they are congested (up to a certain degree), thus sometimes
losing optimality.

\begin{table}[htb]
\begin{center}
\begin{tabular}{l||r|r||}
\hline
   & {\bf Dijkstra's algorithm} & {\bf HH algorithm} \\ \hline
 \# settled nodes & 2275563 & 18966	\\
 \# explored nodes & 2587112 & 36200 \\
 query time [sec] & 11.830 & 0.099 \\
 \hline 
\end{tabular}
\caption{Computational results on the static graph: average values}
\label{tabcomp}
\end{center}
\end{table}

\begin{table}[htb]
\begin{center}
\begin{tabular}{l||r|r|r||}
\hline
   & {\bf Dijkstra's algorithm} & {\bf HH algorithm} & {\bf HH algorithm} \\
   & & {\it naive approach} & {\it heuristic version} \\ \hline
 \# settled nodes & 2280872 & 19174 & 19099	\\
 \# explored nodes & 2594361 & 36581 & 36492 \\
 query time [sec] & 11.917 & 0.100 & 0.099 \\
 distance from optimum (variance) & 0\% & 2.00\% (5.00)& 0.55\% (0.45) \\
 \hline 
\end{tabular}
\caption{Computational results on the graph with dynamic times: average values}
\label{tabcomp2}
\end{center}
\end{table}

\begin{figure}[!ht]
\begin{center}
\fbox{\includegraphics[width=13cm]{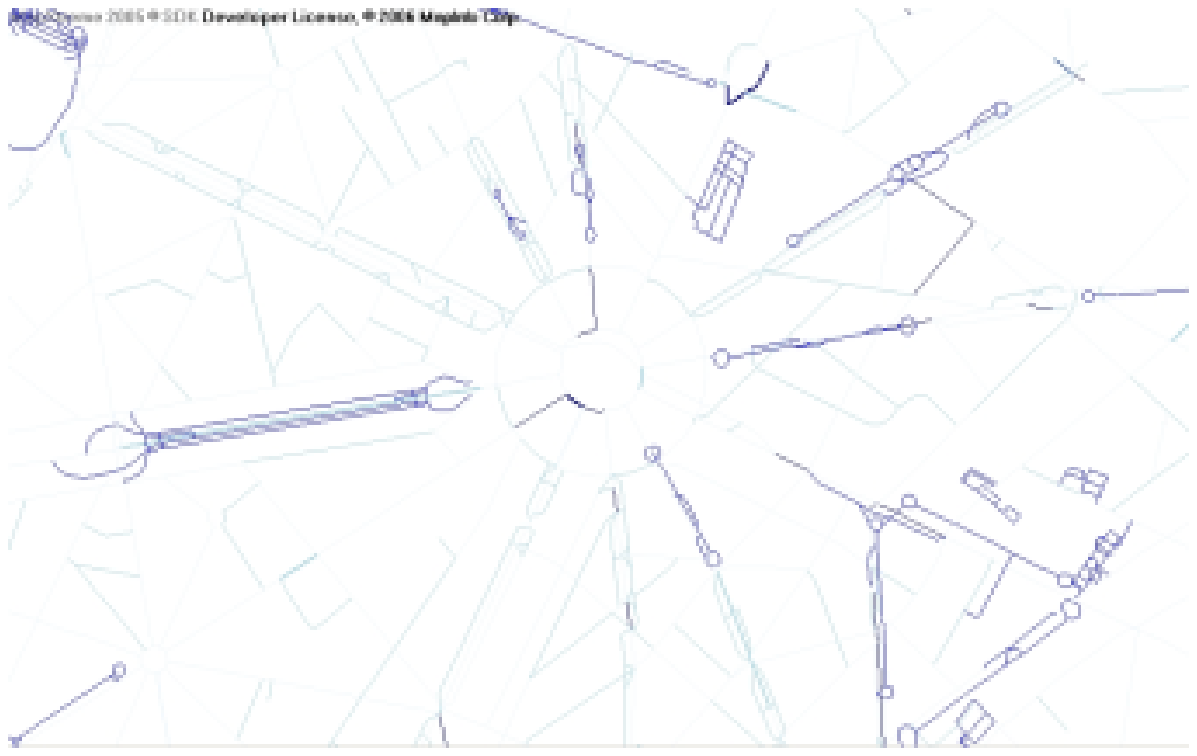}}
\end{center}
\caption{Highway hierarchy near the Champs Elys\'ees, Paris; colour
  intensity and line width increase with hierarchy level, in that
  order (a wide line with a lighter colour has a higher hierarchy
  level than a thin line with dark colour).}
\label{champselyses}
\end{figure}

\section{Conclusion}
\label{conclusion}
We present a heuristic algorithm for efficiently finding fast paths in
large-scale partially dynamically weighted road networks, and
benchmark its application on real-world data. The proposed solution is
based on fast multi-level bidirectional Dijkstra queries on a highway
hierarchy built on the statically weighted version of the network
using the Highway Hierarchies algorithm, and deployed using the dynamic arc
weights. 
Computational results show that, although with no guarantee of optimality,
the proposed solution works well in practice, computing near-optimal
fast paths quickly enough for our purposes.

\section*{Acknowledgements}
We are grateful to Ms.~Annabel Chevaux, Mr.~Benjamin Simon and
Mr.~Benjamin Becquet for
invaluable practical help with Oracle and the real data, and to the
rest of the Mediamobile's energetic and youthful staff for being
always friendly and helpful.

\newcommand{\etalchar}[1]{$^{#1}$}

\end{document}